# Empirical Estimate of the Shape of the Upstream Heliopause
# from IBEX-Lo Helium Measurements: Preliminary Results

Philip A. Isenberg[1], Harald Kucharek[1], and Jeewoo Park[2]

[1] Space Science Center and Department of Physics, University of New Hampshire, Morse Hall, 8 College Road, Durham, NH 03824, USA; phil.isenberg@unh.edu

[2] NASA Postdoctoral Program Fellow, Goddard Space Flight Center, Greenbelt, MD 20771, USA

## ABSTRACT

We present a simplified model of the outer heliosheath to help interpret the observations of interstellar neutral helium by the IBEX-Lo instruments. We assume that the measured particles are composed of the superposition of a primary beam population, with the properties of the local interstellar medium, and a secondary population, created by charge exchange between the primary beam neutrals and the ions that have been deflected as they approach the heliopause. We extract information on the large-scale shape of the heliopause by comparing the helium flux measured at IBEX along four different look directions with simple models of deflected plasma flow around hypothetical obstacles of different aspect ratios to the flow. As a first step in this paper, we model the deflected plasma flow with the analytical solutions for compressible gas flow around a series of oblate ellipsoidal obstacles. Our comparisons between the model results and the observations indicate that the heliopause is very blunt in the vicinity of the heliospheric nose, especially compared to a Rankine half-body or cometary shape. The upstream heliopause seems to be highly elongated in the directions parallel to the interstellar magnetic field, and relatively more compact and symmetric in the directions transverse to that field. The IBEX-Lo helium observations are not consistent with a heliopause elongated in directions parallel to the solar rotation axis.

## 1. Introduction

Studies of the interaction of our heliosphere with the local interstellar medium encounter the fundamental question of the shape of the heliopause boundary. This boundary is the three-dimensional surface that separates the plasma and magnetic field originating at the Sun from that of the surrounding interstellar space. Since its location and shape depend on the properties of the plasmas and fields on both sides, the heliopause represents a sensitive indicator of these properties.

Theoretically, the problem has been effectively circumscribed by Parker (1961). In this work, he assumed a spherically symmetric, shocked, subsonic solar wind providing the interior pressure, and considered the possible extremes of conditions in the external local interstellar medium (LISM). If the dynamical pressure of a subsonic plasma flow around the heliosphere were to dominate the external pressure of the interstellar magnetic field (ISMF), the heliopause would be quasi-spherical on the upwind side with a long cylindrical tail extending downwind, similar to the shape of comets or to the Earth's magnetopause. In the other extreme of dominant magnetic pressure, the internal solar wind would inflate a bubble within the otherwise straight field lines of the ISMF, and the resulting heliopause would be quasi-spherical in directions transverse to the ISMF, with an extended outflow through channels parallel to the field. The former case is equivalent to the hydrodynamic shape known as a Rankine half-body (Batchelor 1967), and its analytical simplicity has lead to its use in many models of the interaction of the solar wind with the interstellar medium (Suess & Nerney 1990; Nerney et al. 1993, 1995; Röken et al. 2015; Isenberg et al. 2015; Sylla & Fichtner 2015; Fahr et al. 2016; Kleimann et al. 2017). The extreme high-field case has also been used to model the ISMF in several recent works (Gamayunov et al. 2010, 2017; Grygorczuk et al. 2014).

Of course, the real heliospheric structure is expected to lie somewhere between these two extremes, as schematically illustrated in Figure 4 of McComas et al. (2009). Our current understanding of this structure depends on a variety of global simulations that make different choices as to the properties of the fluids or kinetic particle populations, the specific physical details to be included, and the numerical schemes being employed. These simulations have been continuously improved as computational resources increase and as spacecraft observations provide more information (Baranov &

Malama 1993; Linde et al. 1998; Zank 1999; Pogorelov et al. 2008; Pogorelov et al. 2013; Borovikov et al. 2011; Heerikhuisen et al. 2014; Izmodenov & Alexashov 2015; Opher et al. 2015). However, many aspects of the heliopause structure are still weakly constrained. The actual position of the heliopause is only known at a single point in space and in time, determined at the Voyager 1 crossing (Gurnett et al. 2013; Krimigis et al. 2013; Stone et al. 2013). We have considerable knowledge of the neutral components of the interstellar medium, including flow direction, coupled speed and temperature, and composition, from the *Ulysses* and *Interstellar Boundary Explorer* (IBEX) missions, but the plasma behavior is less well defined. The undistorted direction of the ISMF is likely given by the consistent indications of the IBEX ribbon center and the deflection of secondary neutrals (Lallement et al. 2010; Funsten et al. 2013; Bzowski et al. 2015; Zirnstein et al. 2016; Kubiak et al. 2016), but major uncertainties persist concerning, for instance, the relation of this field to the Voyager in situ measurements (Burlaga et al. 2013; Burlaga & Ness 2014a, b).

In the light of the complexity of these global simulations and the large range of acceptable parameter values, additional observational constraints would provide an enormous benefit to the modeling effort and to our understanding of the large-scale system. In this paper, we attempt to derive new global constraints from the IBEX data.

The IBEX mission is dedicated to imaging energetic neutral particles from the boundaries of the heliosphere. The IBEX spacecraft was launched in October 2008, eventually attaining a highly elliptical Earth orbit. We will be addressing the observations from the IBEX-Lo instruments, described in Fuselier et al. (2009b). The power of a neutral-particle detector is that remote images can be constructed from the unscattered atoms emitted by a distant large-scale source. In the case of the IBEX mission, a global picture of the boundary of the heliosphere and the regions beyond is available from the distributed neutral fluxes as functions of energy and arrival direction. This imaging capability was central in the detection of the ribbon of energetic hydrogen that surrounds our heliosphere (Fuselier et al. 2009a; McComas et al. 2009, 2014). The sky maps recently presented by Park et al. (2015, 2016) contain uniquely accessible information on the flow of plasma around the heliopause. The purpose of this paper is to extract this information without resorting to complex global simulations. To this end, we will make a number of substantial simplifications that we claim are appropriate to the moderate level of detail contained in the sky maps.

The IBEX instruments can detect fluxes of several neutral species. We choose to work with the helium measurements, since helium is a major constituent of the interstellar medium and since its ionization and recombination characteristics are expected to be simpler than those of hydrogen or oxygen. Interstellar He is ionized primarily by charge exchange with its own ion: $He + He^+ \rightarrow He^+ + He$. This is true for hydrogen as well, but the charge exchange rate for protons and hydrogen is large enough that a significant fraction of the interstellar hydrogen reaching 1 AU has undergone multiple charge exchanges. The nonlinear effects of these multiple encounters means that interstellar hydrogen is strongly "filtered" by the time it reaches the inner heliosphere, resulting in the creation of a "hydrogen wall" upwind of the heliospheric nose (Wallis 1975; Pauls et al. 1995; Baranov & Malama 1993; Katushkina et al. 2016) as well as substantial influence of the shocked solar wind in the inner heliosheath (Desai et al. 2014). In addition, hydrogen undergoes a more complicated passage through the supersonic solar wind, being highly depleted by charge exchange and photoionization, and being further impeded by radiation pressure. Oxygen charge exchanges efficiently with protons, so this species is also nonlinearly affected. In contrast, helium is harder to ionize and these atoms survive the passage from the outer heliosheath with only small modifications to their distribution. In the heliosheath, the effect of multiple charge exchange sources for the helium measured at IBEX can be neglected to the level of accuracy we deal with here.

Helium is also a distinct species in that the IBEX instruments do not detect it directly. Atoms entering the IBEX-Lo detector first encounter a conversion surface, where they produce negative ions that are then identified by standard time-of-flight methods. Incident atoms of hydrogen, carbon, and oxygen primarily "bounce" off the conversion surface with only a small loss of energy, and take on an electron in the process. Helium, however, does not easily gain a third electron. The ions resulting from a helium impact on the conversion surface are sputtered products within a broad range of energies lower than the incident energy. Calibration measurements with a helium beam are incorporated into a statistical analysis to determine how many of these ions are due to a helium encounter (Park et al. 2016). However, the incident energy of the initial helium atom is not well determined by this analysis. Thus, we treat the count rates at IBEX that can be attributed to encounters with interstellar helium atoms as representative of the energy-integrated fluxes of these particles.

The fact that we have no reliable information on the energy of individual helium atoms means that we will be unable to unfold the three-dimensional structure of the source region from the IBEX-Lo data. To invert the IBEX helium sky maps, we would need to know the trajectory of each particle, which requires knowledge of the particle's incident energy. Even the application of a frame transformation, such as from the reference frame of the Earth as it orbits around the Sun to an inertial frame where the Sun is at rest, is beyond the capabilities of this data set. Thus, we are restricted to forward modeling to interpret the helium measurements. In this paper, we will construct trial heliosheaths from our idealized model and derive the resulting energy-integrated fluxes in the reference frame of the Earth in its orbit. We will compare these model fluxes with the observations, and draw conclusions on the plasma flow in the heliosheath and the shape of the heliopause that causes this flow.

We focus specifically on the IBEX-Lo helium maps from Park et al. (2016), taken during the spring when the Earth-orbiting spacecraft is moving into the interstellar flow. This data combines the spring observations from years 2009, 2010, and 2011 (map1 + map3 + map5). In this paper, we will also sum the fluxes in the first three IBEX-Lo energy bins (E-step 1 + 2 + 3), corresponding to the sputtered ions between 11 eV and 77 eV (Galli et al. 2014), to represent the energy-integrated He flux at IBEX. Figure 1 shows the He sky map of the sum of those three energy bins over those three years. This map shows the differential flux intensity of He as inferred from the measured counts and the instrument calibration.

In Figure 2a, we show the same data as Figure 1, but flattened into a plane representing the longitude and latitude look directions of the IBEX-Lo instrument. We have also rotated the map so the longitude coordinate ($\lambda$) is centered on the position of peak flux. The latitude coordinate ($\alpha$) remains centered on the ecliptic plane. In summing the single-energy-step maps of Park et al. (2016), we obtain an average peak position of $(\lambda_o, \alpha_o) \sim (222°, 4°)$ in J2000 ecliptic coordinates.

We interpret the maps in Figure 1 and Figure 2a as the superposition of an undisturbed primary He beam and an irregular cloud of secondary He produced by charge-exchange interactions between the beam and the He$^+$ plasma (Kubiak et al. 2014; Bzowski et al. 2017). At large distance from the Sun, charge exchange within the partially ionized interstellar He beam simply exchanges an ion and atom with the same average velocity and temperature, so has no observable effect. However, as the interstellar particles approach the heliosphere, the ionized component is deflected around the magnetic obstacle of the heliopause, while the neutral component continues in its original direction. Under these conditions, the neutral product of a charge exchange takes on the properties of the deflected plasma and can then communicate these properties to the position of IBEX.

Thus, the secondary cloud tells us about the deflected plasma, and this plasma carries the imprint of the deflecting obstacle: the heliopause. These secondary particles were observed in the initial IBEX measurements (Bzowski et al. 2012, 2015; Kubiak et al. 2014, 2016; Swaczyna et al. 2015), but they have only been recently interpreted as we do here. Despite the conceptual advantage of images such as Figure 1, much greater precision is available when IBEX data is analyzed as a function of the Earth's ecliptic longitude, using the varying position of the spacecraft around the Sun as an energy spectrograph (Möbius et al. 2009). This technique has lead to determinations of the undisturbed neutral gas velocity vector and temperature with unprecedented accuracy (Lee et al. 2012; Möbius et al. 2012, 2015; Leonard et al. 2015). The lopsided extension of the primary beam was also identified in this form of the data, and it was initially analyzed in terms of a separate "warm breeze" component of the interstellar medium (Bzowski et al. 2015; Kubiak et al. 2014, 2016). However, these studies are also consistent with a secondary source for this population (Kubiak et al. 2014), and this interpretation has recently been verified by kinetic simulations of the charge exchange effects (Bzowski et al. 2017).

## 2. A simple model of the plasma in the outer heliosheath

*2.1 Overview*

As stated above, we will construct a deliberately simple model of the nearby interstellar plasma in order to extract qualitative information on the shape of the heliopause from the IBEX He sky maps. We emphasize here that an overly elaborate model, even if it incorporates better physics, can be counter-productive for this purpose when it addresses a level of detail not available from the data. Such elaborations increase the complexity of the calculations and, unless the model results accurately match the observations, one then has to perform the model computations over again with adjusted parameter values. This is not a viable method for interpreting data when each model computation can take several weeks.

Consequently, we propose several substantial simplifications of the system in question. First, we reduce the problem of modeling an irregular shape by taking cuts through the data in characteristic directions, and comparing the fluxes along these cuts with predictions from different sample axisymmetric systems. Specifically, we will construct a series of plasma flow models, using oblate ellipsoidal obstacles of increasing aspect ratio in the flow. Each ellipsoid will have its axis of symmetry aligned with the direction of the undisturbed LISM flow, and will create an axisymmetric pattern of deflected flow depending on the width and curvature of that obstacle. We will then obtain the predicted model He flux along the observational cutlines for each sample obstacle, and compare these model fluxes with the data from those look directions. In this way, we hope to qualitatively characterize the irregular heliopause in terms of its transverse shape in each of the cutline directions.

We choose the cutlines to start at the point of peak flux in the sky map and proceed in straight lines in four directions, as shown in Figure 2a. One of these lines (A) is oriented in the direction where the flux distribution is broadest. One of the other lines (C) is then set anti-parallel to this line, and the other two (B and D) are perpendicular to that direction.

The peak flux position used here is consistent with that determined by Park et al. (2016), in order to interpret their maps without modification. Specifically, we take the peak longitude, $\lambda_o = 222°$, and the peak latitude $\alpha_o = 4°$, in the J2000 coordinate system.

As discussed by Park et al., these values are shifted slightly from the actual primary beam position on the sky as determined by Möbius et al. (2015) due to the energy dependent effects of ionization losses inside the heliopause. We will discuss this issue further in §2.5.

The orientation of these cuts is meant to capture the extremes of the large-scale variations in heliopause shape. We also note that the "maximum" line (A) is necessarily coincident with the direction of the "warm breeze" displacement, since the analysis leading to that interpretation was derived from the same measurements that we use here. The choice is especially meaningful in light of the work by Kubiak et al. (2016), which showed that the warm breeze enhancement lies along the hydrogen deflection plane. This plane also contains the center point of the ribbon circle (Funsten et al. 2013), and is thought to indicate the direction of the undisturbed ISMF.

The second major simplification follows from noting that the irregular shape of the heliopause is itself largely caused by the dynamical effect of the interstellar magnetic field that imposes a preferred direction unaligned with the interstellar flow. In this model, we parameterize this irregular shape in the imposed sample obstacles themselves, understanding that the aspect ratio of the obstacles corresponding to one cutline direction will be different from that representing another direction. Having allowed for a preferred direction in the properties of the obstacle, we then approximate the remaining effect of the ISMF on the plasma flow as contributing to the isotropic pressure forces. Thus, for each sample obstacle, we consider the simplified description of an unmagnetized plasma, characterized by an effective isotropic pressure composed of the sum of the kinetic thermal pressure, $n_{pl} k T_{kin}$ and an isotropic magnetic contribution, $B^2/8\pi$. (Here, $n_{pl}$ is the plasma number density, $T_{kin}$ is an effective isotropic plasma temperature, $k$ is Boltzmann's constant, and $B$ is the amplitude of the ISMF.) We further assume that all the plasma components are co-moving Maxwellians with equal temperatures, and that this plasma evolves adiabatically as it flows around our ellipsoidal obstacles.

As the model plasma is deflected away from the direction of the undisturbed LISM flow, charge exchange between the $He^+$ component of the plasma and the neutral He of the primary beam will create He atoms with the new properties of the deflected plasma. In effect, each volume element of plasma in the OHS will emit He atoms as a consequence of flowing through the undisturbed primary beam. These secondary He atoms are emitted isotropically in the plasma reference frame, at a rate proportional to the densities of the two populations and to the relative speed between them

$$\text{He emission} \ \sim \ \sigma N_{\text{He}} \, n_{\text{He}^+} \left| \mathbf{v}_{pl} - \mathbf{V}_{\text{o}} \right| , \tag{1}$$

where $N_{\text{He}}$ is the undisturbed interstellar neutral density of helium atoms, moving at velocity $\mathbf{V}_{\text{o}}$; $n_{\text{He}^+}$ is the local helium ion density in the deflected plasma which flows at $\mathbf{v}_{pl}$; and $\sigma$ is the charge exchange cross-section, which we take as a constant, $\sigma = 2 \times 10^{-15}$ cm$^2$. The contribution of additional charge-exchange interactions (creating tertiary and higher-order products) is very small, so we assume that the distribution of emitted He corresponds to the Maxwellian distribution of the flowing plasma, ignoring any modifications of the plasma distribution by the addition of distinct pickup ions.

The energy-integrated flux intensity of secondary He which reaches the IBEX detector at a particular look angle, or field-of-view, is then given by the integral of all the emitted atoms from the OHS that have the correct velocity at the point of charge-exchange to end up at the detector. Figure 2b shows the observed fluxes attributed to He, obtained by summing the count rates from IBEX-Lo energy channels 1, 2, and 3, as a function of the angular distance along each cutline in Figure 2a. The experimental and statistical errors on these data points are smaller than the plotted points. We will compare these measurements with the fluxes from our model, using axisymmetric heliopause obstacles of different aspect ratios, to determine the effective ellipticity of the upstream heliopause surface in each of the four directions around the heliospheric nose.

We will find that the regions containing broader distributions of secondary He can be modeled by plasma deflection around broader obstacles. Qualitatively, this is a common-sense, expected result, but here we attempt a more detailed comparison. Our model will incorporate the inherent asymmetry produced by taking the data in the Earth's reference frame, and present results that demonstrate the remaining asymmetry of the plasma flow around the asymmetric heliopause.

In the following subsections, we present the details of our simplified model. We first describe the coordinate system and the specific obstacles used to represent the heliopause in our model. We then present the model solution to the plasma flow around these obstacles, followed by the procedures for calculating the resulting flux of model interstellar He at the position of IBEX.

## 2.2 The coordinate system in the OHS

We construct our model in terms of an axisymmetric oblate spheroidal coordinate system $(\xi, \eta)$, which is one of the standard orthonormal coordinate systems allowing for separable solutions of Laplace's equation. Detailed descriptions of this coordinate system can be found in most texts on mathematical physics. In this paper, the origin of the coordinates is placed at the Sun, and the axis of symmetry, $\hat{\mathbf{z}}$, is taken to point into the undisturbed LISM flow. The ellipsoidal surfaces of constant $\xi$, where $0 < \xi < \infty$, are generated by taking ellipses centered on the origin with foci at $\pm d$ in a direction transverse to $\hat{\mathbf{z}}$, and rotating these ellipses around $\hat{\mathbf{z}}$. The $\xi = 0$ surface is a disc of radius $d$, centered at the Sun and transverse to $\hat{\mathbf{z}}$. Since our model concerns the OHS outside the heliopause, it will not include this singular surface at the Sun. The orthogonal surfaces of constant $\eta$, where $-1 \leq \eta \leq 1$, are hyperboloids of one sheet passing through the $\xi = 0$ disc, and we label the $\eta = 1$ direction as parallel to $\hat{\mathbf{z}}$. The spherical coordinates $(r, \theta)$ with the same origin and symmetry axis are related by

$$r^2 \ = \ \xi^2 + d^2(1 - \eta^2), \qquad r^2 \cos^2\theta \ = \ \xi^2 \eta^2 \ . \tag{2}$$

The third coordinate, corresponding to the azimuthal angle around $\hat{\mathbf{z}}$, can be ignored in our axisymmetric system. The differential scale factors, which appear in the gradient and Laplacian operators, are

$$h_\xi = \sqrt{\frac{\xi^2 + d^2\eta^2}{\xi^2 + d^2}} \qquad \text{and} \qquad h_\eta = \sqrt{\frac{\xi^2 + d^2\eta^2}{1 - \eta^2}} \ . \tag{3}$$

*2.3 The ellipsoidal obstacles*

In this oblate spheroidal coordinate system, the axisymmetric obstacle corresponding to the model heliopause in each case is defined as the $\xi = 1$ surface, and all distances are then measured in terms of the distance along the symmetry axis between the Sun at the origin and the heliospheric nose, $L$. In this paper, we will take the value of this upstream standoff distance to be $L = 120$ AU, which is on the order of the position of the heliopause crossing by Voyager 1.

The different shapes of the sample obstacles will be obtained by changing the value of $d$, with larger $d$ giving broader, blunter heliopause surfaces around the nose. When $d \rightarrow 0$, the oblate spheroid collapses to a sphere centered on the Sun, with $\xi \rightarrow r$ and $\eta \rightarrow \cos\theta$ in that case. In the upstream region, the Rankine half-body of Parker's field-free heliopause roughly corresponds to an ellipsoid with $d = 1$, in that both its curvature at the nose and its transverse extent in the $z = 0$ plane are the same as those of the $d = 1$ ellipsoid. The detailed shapes of the two objects are, of course, different.

Formally, the entire ellipsoidal obstacle in our model has a size transverse to the undisturbed flow equal to $\sqrt{1 + d^2}$ in units of the upwind standoff distance. However, our model focuses on the plasma conditions in the region of the nose, where most of the observed He particles come from, and will not treat any aspects of the OHS beyond about 60° from the LISM inflow axis, $\hat{\mathbf{z}}$. So, even though the application of our model to the observations will find effects corresponding to fairly large values of $d$, this does not imply that we predict an enormous heliosphere. Instead, we will interpret these values in terms of increasingly flat, or blunt, heliopause surfaces in those directions near the nose.

*2.4 The model plasma*

To evaluate the model fluxes incident on the IBEX detector, we need the velocity, density and temperature of the $He^+$ plasma as functions of position in the OHS. In keeping with our deliberately simple model, we represent the undisturbed plasma far upstream of the heliosphere as a uniform isotropic Maxwellian, streaming with velocity $-V_o\,\hat{\mathbf{z}}$, with an effective temperature corresponding to the fast magnetosonic speed, $c_f$, where $c_f{}^2 = c_s{}^2 + V_A{}^2$. Here, $c_s$ is the kinetic sound speed in the plasma, $c_s{}^2 = \gamma k$ $T_{kin}/M_{eff}$; $V_A$ is the Alfvén speed; $\gamma$ is the adiabatic coefficient, $\gamma = 5/3$; and the mass/particle, $M_{eff}$, used to obtain the sound and Alfvén speeds includes both protons and He$^+$ ions. The upstream densities, speed, and temperature are set at the current estimates for these quantities, given in Table 1. We see that the additional isotropic contribution of magnetic forces to our model plasma, represented by $c_f$, results in an effectively subsonic flow of the LISM in the region far upstream, $V_o < c_f$ (McComas et al. 2012).

Within our model axisymmetric system, we obtain the analytical solution for the compressible flow field, $\mathbf{v}_{pl}(\mathbf{x})$, around the heliopause obstacle as a function of the different obstacle aspect ratios, $d$. We start by noting that the number flux in such a flow is conserved, $\nabla \bullet (n_{pl} \mathbf{v}_{pl}) = 0$. If the flow remains subsonic, this flux can be written as the gradient of a scalar potential, $\varphi$, which is a solution of Laplace's equation:

$$n_{pl}\,\mathbf{v}_{pl} \;=\; \nabla\,\varphi, \ \ \text{where} \ \ \nabla^2\,\varphi \;=\; 0. \tag{4}$$

In the upstream 2D quadrant of the spheroidal coordinate system, $0 \le \eta \le 1$, we have boundary conditions $\varphi(\eta = 0) = 0$ and $\partial \varphi/\partial \eta\,(\eta = 1) = 0$, given by the symmetry of the model system. At large distance from the obstacle, the uniform flow corresponds to the potential $\varphi = -n_o V_o\, r \cos\theta = -n_o V_o\, \xi\eta$. On the surface of the obstacle, the normal component of the flux must vanish, so $\partial\varphi/\partial\xi\,(\xi = 1) = 0$.

The separable solutions of Laplace's equation in oblate spheroidal coordinates in this space can be represented by a sum of terms made up of the product of Legendre polynomials, $a_n P_n\,(\xi)\,P_n\,(\eta) + b_n Q_n\,(\xi)\,P_n\,(\eta)$ (see e.g. Morse & Feshbach (1953) p. 1293). The inhomogeneous solution satisfying the boundary condition at large $\xi$ is $\varphi^i = -n_o V_o\, \xi\eta = -n_o V_o\, P_1(\xi)\, P_1(\eta)$. From the orthogonality of the Legendre polynomials, the homogeneous solution is then restricted to the n = 1 term, of which only the $Q_1(\xi)\,P_1(\eta)$ component goes to zero at large $\xi$. Thus, the potential solution is simply

$$\varphi = -n_o V_o \Big[ P_1(\xi)P_1(\eta) + A Q_1(\xi)P_1(\eta) \Big]$$

$$= n_o V_o \eta \left[ -\xi + A \left( \frac{\xi}{d} \operatorname{arccot} \left( \frac{\xi}{d} \right) - 1 \right) \right], \tag{5}$$

for real $\xi$. The application of the boundary condition at $\xi = 1$ sets the value of $A$ as

$$A = d \left[ \arctan(d) - d / (1 + d^2) \right]^{-1} \tag{6}$$

where $d$ is measured in units of the heliopause standoff distance $L$.

We take the overall plasma to behave adiabatically, so the pressure and the density are related by $n_{pl} k T_{eff} \sim n_{pl}^{5/3}$, or

$$n_{pl} c_f^{-3} = N_c = \text{constant}. \tag{7}$$

Invoking Bernoulli's law, the effective sound speed in the model flow behaves as

$$c_f^2 = c_o^2 - \frac{\gamma - 1}{2} v_{pl}^2, \tag{8}$$

where the constant $c_o$ is the value of this effective speed at the flow stagnation point, $v_{pl} = 0$, found at the nose of the heliopause obstacle $(\xi, \eta) = (1, 1)$. Using the upstream parameter values from Table 1, we find that $c_o = 44.0$ km s$^{-1}$.

Now, we multiply (8) by $n_{pl}^2$ and recognize that $(n_{pl} v_{pl})^2$ is given at all points in the OHS by $\nabla \varphi \cdot \nabla \varphi \equiv W$, which is known through (5). Using (7), we find that $c_f^2$ is given by one of the roots of the quartic,

$$c_f^8 - c_o^2 c_f^6 + W/(3N_c^2) = 0. \tag{9}$$

When $W = 0$, the relevant root is $c_f = c_o$, and this branch continues to be the largest real root of (9) until $W$ reaches a critical value of

$$W_{crit} = \left( \frac{81}{64} - \frac{273}{256} \right) N_c^2$$

at $c_f^2 = 3 c_o^2/4$. At this value, all the roots of (9) become complex and the continuous, adiabatic solution of Laplace's equation is no longer valid for the assumed upstream conditions. In physical terms, this critical point occurs if the speed of the model plasma, as it accelerates around the obstacle, reaches its effective sound speed. Continuous flow of a real plasma under these circumstances must be described by supersonic equations, and might require a shock to be formed at these positions. However, this is not a serious issue for our simple model since the critical points are not reached until the plasma has traveled well around the flanks of the obstacles. The positions of these points depend on $d$, the aspect ratio of the particular sample obstacle, but they are all far from the nose region where we wish to compare with the IBEX data. We will not attempt to apply our simple model in the effectively supersonic regions beyond the critical points.

This derivation of the compressible flow around the heliopause is similar to the recent analysis presented by Kleimann et al. (2017) of the fields and flows around a Rankine half-body. We note, however, that Kleimann et al. claim their model can be usefully extended into the supersonic regions around the flanks of their heliopause. We make no such claim here, but rather treat these regions as inaccessible with our model.

In order to obtain quantitative He fluxes from this model, we must specify the portion of the model plasma that corresponds to $He^+$. We have assumed that all the plasma components flow together at the same temperature, so the helium ion velocity is just $\mathbf{v}_{pl}$ and the fractional number density is constant at $n_{He^+} = 0.1\, n_p$, as in Table 1. In addition, though, the total effective temperature must be partitioned between the thermal contribution of the particles and the modeled effective contribution of the magnetic field. We define a partition function, $\zeta$, to denote the fraction of the effective temperature attributable to the thermal motion of the ions, $T_{He^+} = \zeta\, T_{eff}$. The simplest assumption for $\zeta$ would be to take it constant, equal to the upstream value of $(c_s\, /c_f)^2$. However, this would underestimate the plasma compression and heating that occurs near the nose, where many of the detected atoms originate. Intuitively, the partition function should depend on the plasma density, such that the thermal fraction increases as the plasma slows and compresses. Likewise, as the plasma accelerates and becomes less dense, one would expect the effective magnetic field pressure to take on a relatively greater fraction of the total. For guidance on this point, we refer to the work of Izmodenov & Alexashov (2015, hereinafter I&A) who reported on their 3D kinetic-MHD heliospheric model, and presented plots of many plasma and field quantities along the stagnation line in their

Figure 4. Figure 4A of I&A shows the plasma density and Figure 4B shows both the thermal pressure and the total (magnetic + thermal) pressure from their global model. The I&A increase in plasma density from their upstream boundary to the nose of the heliopause is about 20%, which is comparable to that in our model compressible flow. Over that same distance, the I&A pressure plot shows that the partition of thermal to total pressure increases by a factor of ~ 5 close to the stagnation point.[†] To incorporate a qualitatively similar thermal behavior in the simplest possible way, we take the partition function in our model to be defined by two straight lines, one connecting the undisturbed upstream conditions with those at the compressed stagnation point, and the other connecting the upstream conditions with a zero-density fluid dominated by magnetic pressure:

$$\zeta(n_{pl}) = \begin{cases} 19.78\, n_{pl} - 0.826 & n_{pl} \geq 0.044 \\ \\ n_{pl} & n_{pl} < 0.044 \end{cases}, \qquad (10)$$

where the total ion density, $n_{pl}$, is measured in $\text{cm}^{-3}$. Within our model, the total ion density remains in the range $0.03 < n_{pl} < 0.053$ $\text{cm}^{-3}$, so this partition function is only applied for values of $0.03 < \zeta < 0.22$, far from the extreme limits.

*2.5  The model fluxes*

The plasma described in the previous section will flow toward the heliopause from the direction of the primary He beam and be deflected around the sample ellipsoidal obstacles of various aspect ratios placed in its path. For each value of the aspect ratio, parameterized by $d$, our simple model specifies the density, velocity and temperature of the He$^+$ component everywhere in the subsonic region beyond the heliopause. This information allows us to compute the sample primary and secondary He fluxes at IBEX, for comparison with the sky maps of Park et al. (2016).

---

[†] We note that these rough estimates deliberately neglect small features in the I&A plots which may be indicative of a plasma depletion layer just outside the heliopause (Cairns & Fuselier 2017). We do not expect such relatively localized structures to be evident in the large-scale moderate-resolution He maps, so there would be no point in trying to include them here.

We take the IBEX spacecraft to orbit the Sun at the Earth's orbital speed of $u_E =$ 29.8 km/s (neglecting the small additional velocity of the spacecraft's motion around the Earth), and we take this orbit to be a circle of radius 1 AU in the ecliptic plane. Atoms enter the instrument apertures transverse to the Earth-Sun line, so they are detected essentially at the perihelion of their motion through the Sun's gravitational field (Möbius et al. 2012). The sky maps are constructed by equating the look direction tangent to the Earth's orbit on a given day of year with a longitudinal value, $\lambda$, and equating the look direction during the spacecraft spin (about the axis pointing toward the Sun) with a latitude value, $\alpha$. The flux intensities measured at these times are then plotted at the corresponding angles in the J2000 coordinate system. However, atoms at different incident energies will have traveled along different trajectories from different regions of space to reach the spacecraft. Since we do not have accurate information on the energy of the incident He atoms, we must resort to accumulating these particles only as a function of this latitude and longitude at the spacecraft, even though they come from different spatial regions.

For each longitudinal position of IBEX, and each latitudinal look angle, the trajectories of incident He atoms are simple functions of $u_i$, their speed at IBEX in the inertial frame of the Sun. The particles follow the standard trajectories of test particles in a gravitational system centered on the Sun. Each trajectory is confined to a plane containing the Sun and the IBEX position, and is given by

$$r = \frac{l^2}{k_g(1 + \varepsilon \cos \omega)},$$ (11)

where $(r, \omega)$ are the radial and angular coordinates in the plane of motion, measuring $\omega$ from the perihelion, and $k_g$ is the Sun's gravitational constant, $k_g = 887$ (km/s)$^2$/AU. The trajectory of each particle conserves the particle's total energy/mass, $E = v^2/2 - k_g/r$, its angular momentum/mass, $l = r v_\omega$, and the ellipticity,

$$\varepsilon = \sqrt{1 + \frac{2El^2}{k_g^2}} \,.$$ (12)

Taking the perihelion at $r = 1$ AU and specifying $u_i$ sets all these constants and determines the trajectory. This trajectory will be a closed ellipse ($\varepsilon < 1$) for small $u_i$ or an open hyperbola ($\varepsilon > 1$) for larger $u_i$, with the discriminating value of $u_i{}^2 = 2\,k_g \times 1$ AU ($u_i \sim 42.12$ km/s) resulting in a parabolic trajectory.

For a given IBEX position, we will only consider the "direct" trajectories from the upwind region of the OHS, and ignore the possible "indirect" particles that can loop around the Sun before arriving at IBEX. As they approach the Sun, interstellar atoms are subject to additional ionization by charge exchange and photoionization, and the longer trajectories of the indirect particles near the Sun causes the contribution of these particles to the measured flux to be small.

Furthermore, following Park et al. (2016), we will not include the effects of this additional ionization inside the heliopause in this analysis. The main effect of these processes on the direct particles is an overall depletion of the intensity, on the order of 20%. Since our model comparisons will only use normalized fluxes, such an overall depletion does not impact our results. A smaller effect is due to the energy dependence of the ionization, which favors the survival of the higher-energy neutrals. This dependence leads to a shift of the apparent peak flux location of several degrees, as mentioned in §2.1 and detailed in Park et al. (2016). In this way, we can compare our model results directly with the fluxes presented in Park et al.

In the IBEX reference frame, the incident particle speed will be

$$u = u_E \cos\alpha + \sqrt{u_i{}^2 - u_E{}^2 \sin^2\alpha}\,, \tag{13}$$

where $\alpha$ is the latitudinal look direction. The energy-integrated flux in this look direction, at longitudinal position $\lambda$, will then be

$$j(\lambda,\alpha) = \frac{1}{4\pi}\int u^3\,du\,F(\lambda,\alpha,u) \tag{14}$$

where $F$ is the density of He atoms arriving at IBEX with the designated speed and direction. The secondary component of this flux is given by accumulating the charge-exchanged neutral particles that are emitted in the OHS at the rate (1) at the particular position and velocity to reach IBEX with the designated speed and look direction.

Denoting a position in our model OHS by **x** and the individual He velocity along its trajectory by $\mathbf{v}_t(\mathbf{x})$, we write the secondary component of the incident density as

$$F_s = \sigma N_{He} \int dt\, n_{He^+} \left| \mathbf{v}_t - \mathbf{V}_o + \sqrt{\frac{8kT_o}{\pi M}} \right| \left( \frac{M}{2\pi kT_{pl}} \right)^{3/2} \exp\left[ -\frac{M}{2kT_{pl}} \left( \mathbf{v}_t - \mathbf{v}_{pl} \right)^2 - \frac{t}{\tau} \right] \quad (15)$$

where the integral over $dt = ds/\upsilon_t(\mathbf{x})$ is taken along the particle trajectory that ends at IBEX with the appropriate speed and direction. We take the He$^+$ plasma to have a Maxwellian distribution with density $n_{He^+}(\mathbf{x})$, flow velocity $\mathbf{v}_{pl}(\mathbf{x})$, and temperature $T_{pl}(\mathbf{x})$ given by the model of §2.4. We also approximate the velocity of the primary neutral beam in the charge-exchange rate (within the absolute value term) as the bulk velocity $\mathbf{V}_o$ plus a contribution from the thermal motion within the beam, taken as the most-probable-speed of a Maxwellian, $\upsilon^2_{mps} = 8kT_o/\pi M$ (Bzowski et al. 2012). The integral along each trajectory is limited by the possibility of re-ionization of the atom by charge exchange before the particle arrives at IBEX, and for this tertiary process we approximate the time scale by the constant value from the undisturbed system, $\tau = \sigma\, n_{He^+}\, V_o = 4.8 \times 10^{10}$ s.

In practice, we start the computation of (14) at the value of $u_i$ required to reach the model heliopause in the given direction, $u_i > 41.9$ km/s, and increment the incident speed to larger values. For the closed trajectories, we perform the integral (15) in both directions inward from the aphelion of the trajectory to the heliopause and sum the results. For the open trajectories, we start at the heliopause, integrate outward to a maximum distance of 400 $L$, and then add a contribution from the primary beam

$$F_p = N_{He} \left( \frac{M}{2\pi kT_o} \right)^{3/2} \exp\left[ -\frac{M}{2kT_o} \left( \mathbf{v}_t - \mathbf{V}_o \right)^2 \right]. \quad (16)$$

Since the density of He$^+$ inside the heliopause is small, we ignore further modifications of the He population from charge exchange over the portion of the trajectory within the ellipsoidal obstacle. We also do not explicitly include the depletion of the neutral He due to photoionization close to the Sun. This depletion is an essentially constant effect, which disappears from our normalized model fluxes.

## 3. Results

In this section, we present model results of He energy-integrated fluxes in the reference frame of the IBEX spacecraft, using different values of the aspect ratio parameter, $d$, for the sample ellipsoidal obstacles representing the heliopause. Our goal is to obtain an empirical estimate of the large-scale asymmetry of the upstream heliopause by comparing these results with the fluxes measured along cuts through the He sky map in different directions from the position of the peak flux.

A determination of the heliopause asymmetry is complicated, however, by the fact that the measured fluxes are inherently asymmetric due to their observation in the IBEX reference frame orbiting the Sun. The Earth's orbital speed is comparable to the particle speeds, and so distorts the apparent arrival direction of the particles. This effect was recognized in all the papers that analyzed the primary beam (Lee et al. 2012, 2015; Möbius et al. 2012), and we find comparable results here.

Figure 3 shows the sky map of He flux from our model for the $d = 5$ case, which illustrates the appearance of a sky map created from a sample fully axisymmetric source. The display shows the model flux on a $2° \times 2°$ grid, normalized to the value at the peak. The origin position is set as in Figure 2a, so the 0° longitude ($\lambda$) has been shifted to the peak position, but the latitude ($\alpha$) remains relative to the ecliptic plane. The black areas in this figure correspond to look directions where at least one of the computed particle trajectories encounters the supersonic region where our model does not apply.

In this figure, one can clearly see a broadening in longitude, along with more latitudinal spreading on the right (starboard) of the peak compared to the left (port) side. Thus, a portion of the asymmetry seen in the sky map of Figure 2 is simply a result of measuring the fluxes in the Earth's reference frame. This inherent distortion is included in our model, so the individual flux curves from the sample obstacles all contain this effect. Thus, the differences between the model fluxes in the cases shown below correspond to the effect of different obstacle shapes, unfolded from the distortion of the Earth's motion.

In Figures 4 and 5, we show comparisons between the data and our model along the cutlines of Figure 2, as functions of the angular distance along the line from the position of the peak flux. The total measured count rates in the IBEX-Lo energy steps 1 - 3 are shown as diamond-shaped points at equally-spaced angular positions, from Figure 2b. The different energy-integrated fluxes from the model are shown as solid curves, normalized to the observed peak flux. Figure 4 combines the two quasi-vertical cutlines of Figure 2 (lines B and D) into a "latitudinal" plot, while Figure 5 combines the two quasi-horizontal cutlines (lines A and C) into a "longitudinal" plot. The model fluxes are calculated using plasma flows around different axisymmetric ellipsoidal heliopause shapes, taking (from lower to upper in the figure) $d$ = 5, 10, 20, 40, and 80. As we emphasized in the last section, we do not apply our model to the flanks of the heliopause, and we do not interpret these values of $d$ to indicate enormous heliospheres thousands of AU wide. Rather, we use these values to indicate the relative degree of curvature or bluntness (lack of curvature) and the heliopause shape near the nose. For comparison, we also show the normalized fluxes of the primary beam alone, setting the secondary contribution to zero, as the dashed curves in the figures.

We first see from these figures that the model curves are not symmetric about the point of peak intensity, even though the plasma flow properties in each case are completely axisymmetric. This is the same point that was illustrated in Figure 3, due to the distorting effect of the Earth's orbital motion. We also see that the ellipsoidal shapes assumed by our model do not yield particularly good fits to the data. In Figure 4, the data points show evidence of increasing curvature relative to the ellipsoids as the distance from the peak increases. While the points at ± 21˚ lie near the $d$ = 80 curve, at ± 27˚ they are nearer to the $d$ = 20 curve, and beyond ± 33˚ they fall to the $d$ = 10 curve or below. In Figure 5, the "longitudinal" points appear above all the model curves and do not follow them at all.

Several conclusions can still be made, however. The data points of Figure 4 indicate that the heliopause is essentially symmetric about the nose in that "latitudinal" direction, since the equally spaced points hit approximately the same model curves at the same distances from the peak. It is difficult to assess the symmetry in the "longitudinal" direction from Figure 5, but it is clear that the heliopause is substantially flatter and broader in this direction, since the data points are all much higher than any of the model curves from the same cases as Figure 4.

The data points also indicate a much broader heliopause than would be obtained from a $d$ = 1 ellipsoid (not shown), which has a curvature about the nose equivalent to that of the Rankine half-body structure (Parker 1961; Isenberg et al. 2015; Röken et al. 2015; Kleimann et al. 2017) . Thus, we conclude that the heliopause structure about the nose is very much blunter than that implied by a Rankine half-body, or any other quasi-spherical shape.

Our choice for the cutline directions of Figure 2 were purposely made to be consistent with the parallel and perpendicular directions relative to the projections of the undisturbed ISMF on the plane of the sky. In that context, the comparisons in Figures 4 and 5 show that the heliopause is greatly extended along the ISMF and more confined in the directions transverse to it. Thus, the empirical evidence from the IBEX He observations suggests that the shape of our heliosphere is closer to that of Parker's (1961) strong magnetic field limit than to the weak field limit represented by the Rankine half-body shape.

We also note that the ISMF is not thought to be transverse to the LISM flow, but is almost diagonally oriented off the starboard bow of the heliosphere. A natural consequence of the heliopause distortion by such a tilted field would be a shift of the heliospheric nose away from its nominal position on the line between the Sun and the LISM inflow (Ratkiewicz et al. 1998; Ratkiewicz et al. 2000). Such a shift was not included in our model, which assumed that all the ellipsoidal obstacles were axisymmetric about the inflow axis. However, the skewed data points in Figure 5, consistent with a distinct warm breeze of He, may instead simply imply a distortion of the heliopause shape due to the tilted magnetic field. Such shifts can also be seen in global MHD simulations (e.g. Opher et al. 2006; Pogorelov et al. 2008; Izmodenov & Alexashov 2015).

On the other hand, these model results and data interpretations are not consistent with the recent claims by Opher and colleagues that the wound-up interplanetary magnetic field connected to the Sun has a dominant effect on the shape of the heliopause and heliosphere (Opher et al. 2015, 2016; Drake et al. 2015). The heliopause shapes predicted by those models is extended in the directions parallel to the solar rotation axis, rather than parallel to the ISMF. The "croissant-shaped" heliopause of Opher et al. would create a secondary He cloud which was primarily extended in ecliptic latitude, contrary to the observations modeled here.

## 4. Summary and Conclusions

The measurement of interstellar neutral helium by the IBEX-Lo instruments reveals an asymmetric distribution of particles as seen on the plane of the sky. We interpret this distribution as the superposition of a primary beam with properties of the LISM, and a cloud of secondary particles resulting from charge exchange of the primary beam atoms with the interstellar plasma as it is deflected around the heliopause.

In this paper, we presented a simplified model of the plasma flow in the outer heliosheath as a means of evaluating the heliopause structure underlying the observations. The model took the heliosheath plasma to behave as a compressible, isotropic, Maxwellian gas, and we derived the flow of that gas around a series of oblate ellipsoidal obstacles meant to represent a range of possible heliopause shapes. Although the specific He fluxes predicted by these model shapes did not match the data very well, several qualitative conclusions can be made.

We find that the heliopause is very blunt near the nose, rather than rounded in a quasi-spherical manner. The heliopause is also much more extended in the directions along the ISMF than in the directions transverse to the field. It appears to be shaped symmetrically in the directions transverse to the ISMF, but may be skewed along the direction of the field due to the tilt of the field with respect to the flow. Finally, we claim that the IBEX He observations are inconsistent with a heliopause oriented along the solar rotation axis, as advocated by Opher et al. (2015, 2016).

*Acknowledgements*. The authors are grateful for valuable conversations with E. Möbius, M. A. Lee, T. G. Forbes, B. J. Vasquez, G. P. Chini, J. P. McHugh, and V. V. Izmodenov. This work was funded in part by NASA grants NNX14AJ53G and NNX07AC14G, and by the Interstellar Boundary Explorer  mission as a part of the NASA Explorer Program. J. Park was also supported by the NASA Postdoctoral Program at the Goddard Space Flight Center, administered by USRA through a contract with NASA.

**Table 1.** Model parameters for the heliosphere and outer heliosheath

| | | |
|---|---|---|
| LISM proton density | $n_p$ | 0.04 cm$^{-3}$ |
| LISM He$^+$ density | $n_{He^+}$ | 0.1 $n_p$ |
| LISM neutral helium density | $N_{He}$ | 0.015 cm$^{-3}$ |
| LISM inflow speed | $V_o$ | 26 km s$^{-1}$ |
| LISM temperature | $T_{kin}$ | 7000 ˚K |
| ISMF intensity | $B_o$ | 4.4 $\mu$G |
| LISM Alfvén speed | $V_A$ | 40.6 km s$^{-1}$ |
| LISM sound speed | $c_s$ | 8.3 km s$^{-1}$ |
| LISM fast-mode speed | $c_f$ | 41.4 km s$^{-1}$ |
| HP stand-off distance | $L$ | 120 AU |
| Solar gravitational constant | $k_g$ | 887 km$^2$ s$^{-2}$ AU$^{-1}$ |
| LISM inflow longitude | $\lambda_o$ | 222° |
| LISM inflow latitude | $\alpha_o$ | 4° |

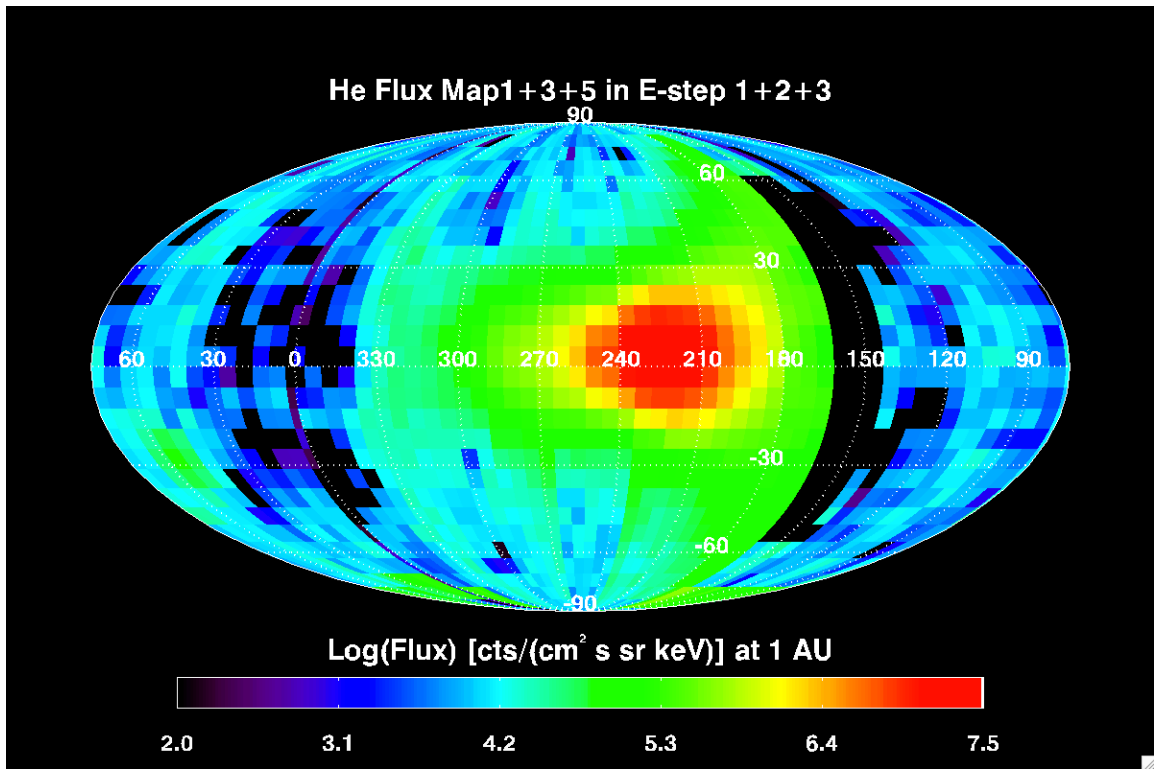

**Figure 1.** Map of He flux incident at IBEX in the energy range 11 - 77 eV (sum of fluxes in E-steps 1, 2, and 3) from the spring observations in 2009, 2010, and 2011 (map1 + map2 + map3). The values are obtained from the analysis of Park et al. (2016), and shown in a Mollweide projection in J2000 ecliptic coordinates.

a)

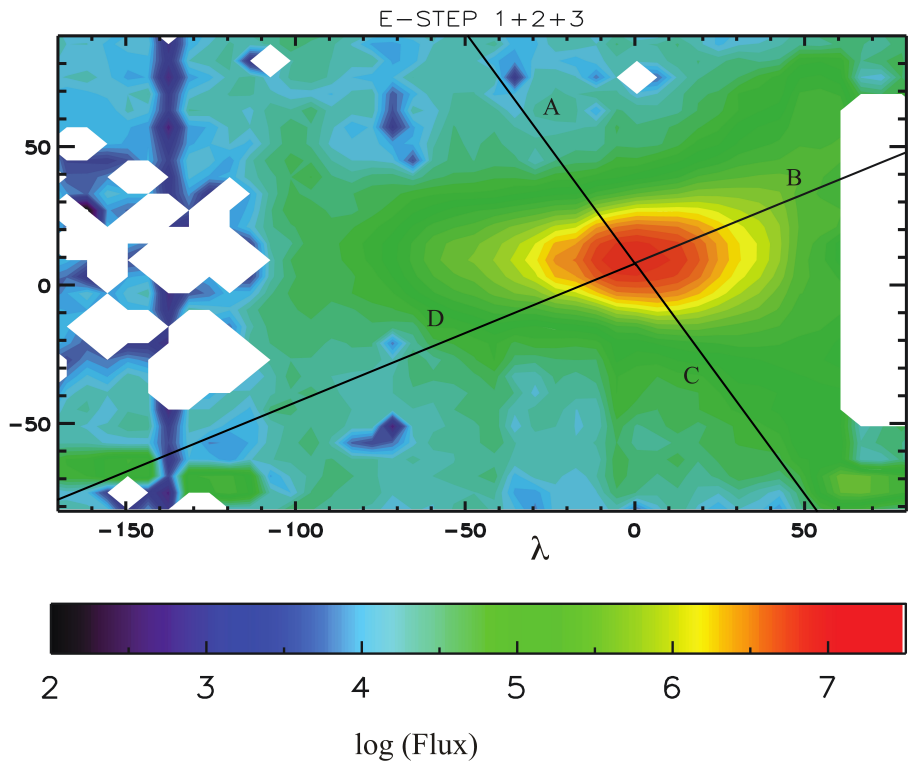

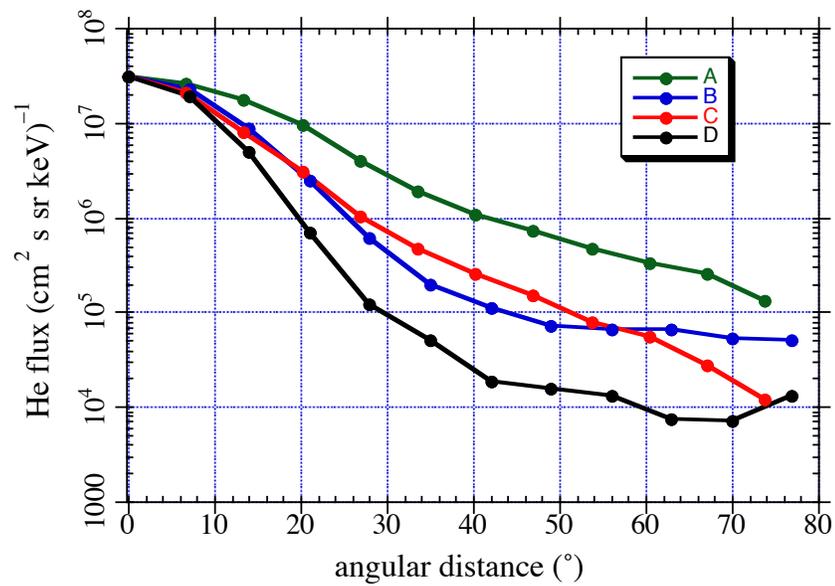

**Figure 2.** a) Portion of the He sky map of Fig. 1, flattened to longitude vs. latitude in J2000 ecliptic coordinates, and rotated in longitude to center on the peak He flux. Four cutlines have been drawn from the peak position. Lines A and C are chosen parallel to the estimated direction of the undisturbed ISMF. Lines B and D are taken perpendicular to the field direction. b) Sum of He flux in the first three energy steps of IBEX-Lo using map1 + map3 + map5 along the cutlines of Fig. 2a.

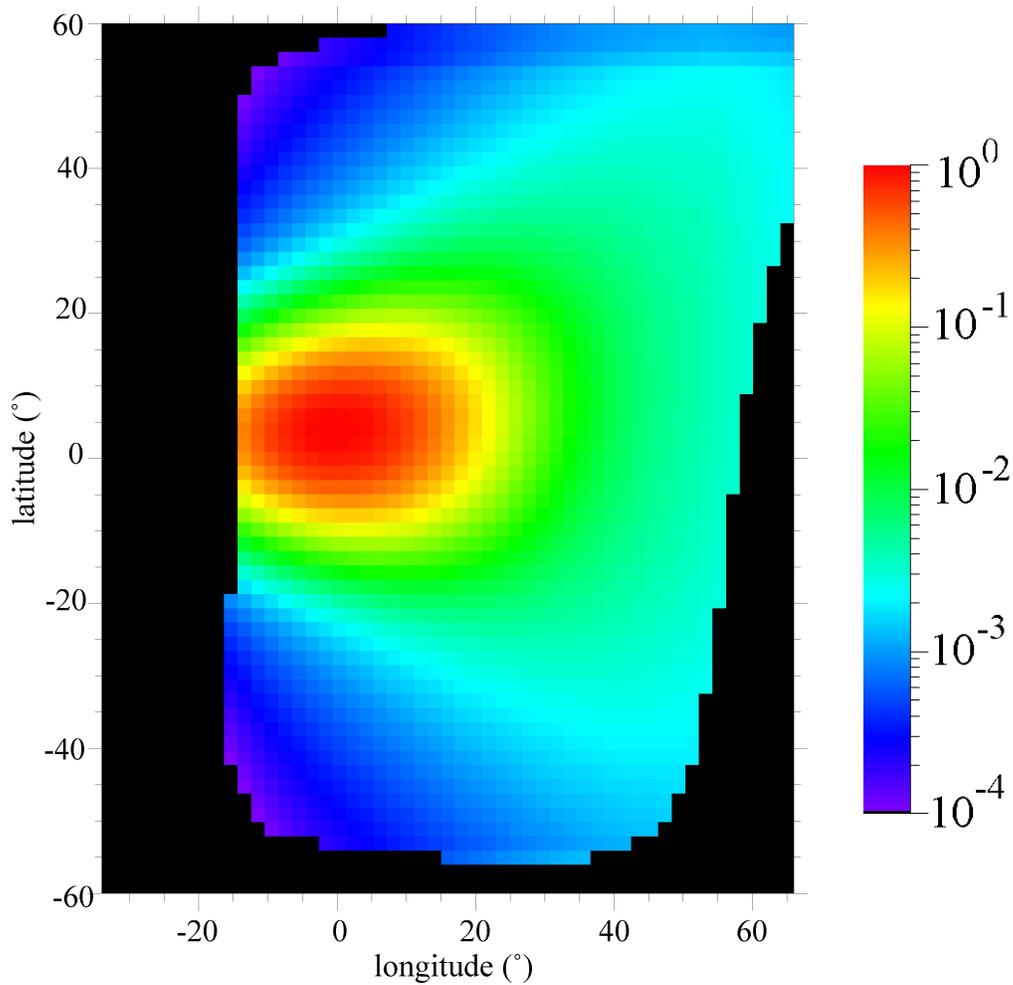

**Figure 3.** Model He flux in the IBEX reference frame, normalized to the peak value, for an ellipsoidal heliopause with $d = 5$. Longitude has been shifted to place the peak flux at zero degrees as in Fig. 2a. Latitude is relative to the ecliptic plane.

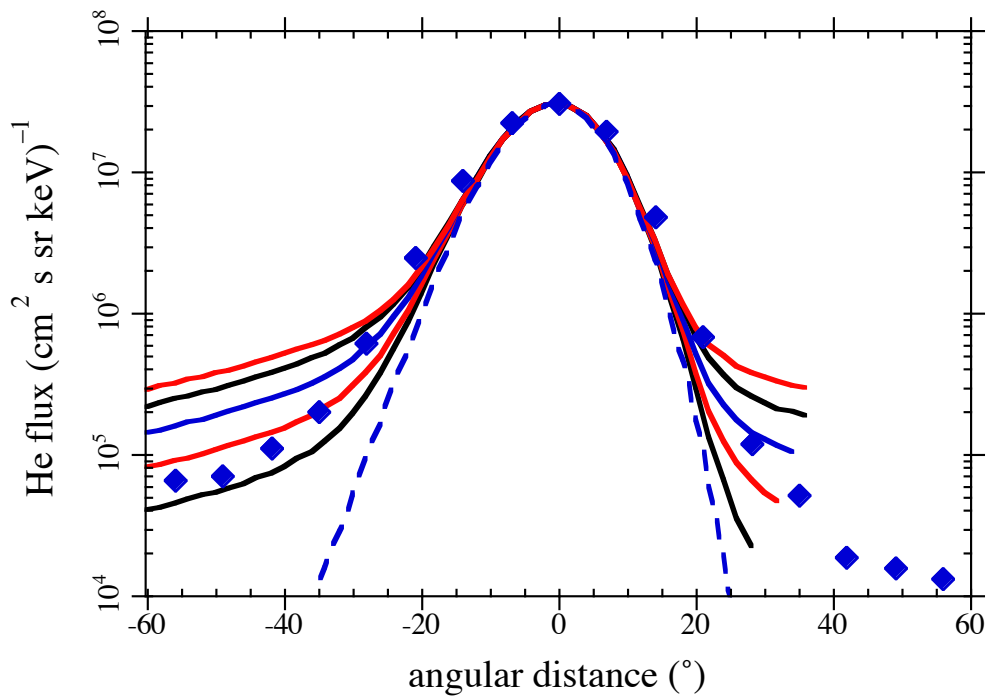

**Figure 4.** Model fluxes and data along cutlines B and D from Fig. 2, normalized to the peak observed flux, as functions of angular distance along those lines. Smooth curves show model fluxes for (from bottom) $d$ = 5, 10, 20, 40, and 80. The dashed line shows the modeled primary beam alone. Blue diamonds show the data points from Fig. 2b.

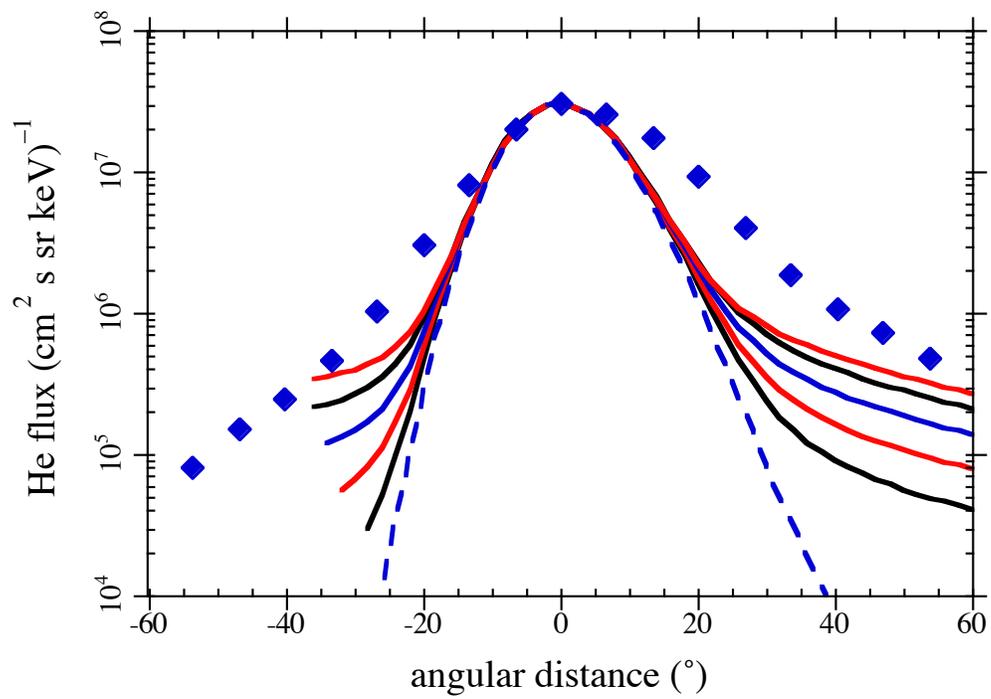

**Figure 5.** As in Fig. 4 for cutlines A and C.